\documentstyle[12pt,epsf,a4]{article}

\def\dfrac#1#2{
  {\displaystyle {#1 \over #2}}}

\def\simge{
  \mathrel{\rlap{\raise 0.511ex \hbox{$>$}}{\lower 0.511ex \hbox{$\sim$}}}}
\def\simle{
  \mathrel{\rlap{\raise 0.511ex \hbox{$<$}}{\lower 0.511ex \hbox{$\sim$}}}}
\def\slashchar#1{
  \setbox0=\hbox{$#1$} \dimen0=\wd0 \setbox1=\hbox{/} \dimen1=\wd1 
  \ifdim\dimen0>\dimen1 \rlap{\hbox to \dimen0{\hfil/\hfil}} #1 
  \else \rlap{\hbox to \dimen1{\hfil$#1$\hfil}} / \fi}

\catcode`\@=11
\def\Let@{\relax\iffalse{\fi\let\\=\cr\iffalse}\fi}
\def\vspace@{\def\vspace##1{\crcr\noalign{\vskip##1\relax}}}
\def\multilimits@{\bgroup\vspace@\Let@
 \baselineskip\fontdimen10 \scriptfont\tw@
 \advance\baselineskip\fontdimen12 \scriptfont\tw@
 \lineskip\thr@@\fontdimen8 \scriptfont\thr@@
 \lineskiplimit\lineskip
 \vbox\bgroup\ialign\bgroup\hfil$\m@th\scriptstyle{##}$\hfil\crcr}
\def\Sb{_\multilimits@}
\def\endSb{\crcr\egroup\egroup\egroup}
\def\Sp{^\multilimits@}

\catcode`\@=\active

\newcommand{\be}{\begin{equation}}
\newcommand{\ee}{\end{equation}}
\newcommand{\bea}{\begin{eqnarray}}
\newcommand{\eea}{\end{eqnarray}}
\newcommand{\nn}{\nonumber}
\newcommand{\lp}{\left(}
\newcommand{\rp}{\right)}
\newcommand{\MSbar}{\overline{\mbox{MS}}}
\newcommand{\RI}{\mbox{RI}}
\newcommand{\as}{\alpha_s}

\begin{document}

\pagestyle{empty}

\begin{flushright}
ROME1-1198/98 \\
\end{flushright}

\vskip 1cm 
\centerline{\LARGE {\bf {\ Quark Mass Renormalization}}}
\vskip 0.1cm  
\centerline{\LARGE{\bf{ in the $\MSbar$ and RI schemes}}} 
\vskip 0.3cm 
\centerline{\LARGE{\bf{ up to the NNLO order }}} 

\vskip 1cm 
\centerline{\bf{ Enrico Franco$^a$ and Vittorio Lubicz$^b$}} 

\vskip 0.5cm 
\centerline{$^a$ Dip. di Fisica, Universit\`a degli Studi di Roma 
``La Sapienza" and} 
\centerline{INFN, Sezione di Roma, P.le A. Moro 2,
00185 Roma, Italy. } 
\centerline{$^b$ Dip. di Fisica, Universit\`a di Roma Tre and
INFN, Sezione di Roma,} 
\centerline{Via della Vasca Navale 84, I-00146 Roma,
Italy} \vskip 1cm

\begin{abstract}
We compute the relation between the quark mass defined in the minimal 
modified $\MSbar$ scheme and the mass defined in the ``Regularization 
Invariant" scheme ($\RI$), up to the NNLO order. The $\RI$ scheme is 
conveniently adopted in lattice QCD calculations of the quark mass, since 
the relevant renormalization constants in this scheme can be evaluated 
in a non-perturbative way. The NNLO contribution to the conversion factor
between the quark mass in the two schemes is found to be large, typically 
of the same order of the NLO correction at a scale $\mu \sim 2$ GeV. 
We also give the NNLO relation between the quark mass in the $\RI$ scheme 
and the renormalization group-invariant mass.
\end{abstract}

\newpage
\pagestyle{plain} \setcounter{page}{1}

\section{Introduction}

The values of quark masses are of great importance in the phenomenology of 
the Standard Model and beyond. For instance, bottom and charm quark masses 
enter significantly in the theoretical expressions of inclusive decay rates 
of heavy mesons, while the strange quark mass plays a crucial role in the 
evaluation of the $K \rightarrow \pi\pi$, $\Delta I=1/2$, decay amplitude and 
of the CP-violation parameter $\epsilon^\prime/\epsilon$. In the Standard 
Model, quark and lepton masses are fundamental parameters. Quark masses, 
however, cannot be measured directly, since quarks do not appear as physical 
states and are confined into hadrons. Thus, for quarks the kinematical concept 
of on-shell mass is meaningless. The values of the quark masses then depend 
on precisely their definitions, which can can be given in terms of a short 
distance mass in some renormalization scheme.

A popular definition of the quark mass is the $\MSbar$ mass. For each flavour 
of quarks, this quantity is a well defined, short distance running coupling,
provided the relevant scale $\mu$ is chosen inside the perturbative region 
of QCD. An alternative, but equally convenient definition of the quark mass 
is the so called invariant quark mass. This has the major advantage of being 
both a scale and a scheme independent quantity. At present, the connection 
between the invariant mass and the $\MSbar$ mass is known up to four-loop 
order in QCD \cite{g4a}. 

For heavy quarks only, the pole mass has also been considered and its 
relation to the $\MSbar$ mass has been computed up to two-loop order 
\cite{pole}. Since, however, for confined quarks there should not be any pole 
in the full propagator, the definition of the pole mass makes sense only in 
perturbation theory. On the other hand, the pole mass is affected by a 
renormalon ambiguity \cite{renorm1,renorm2}, which prevent the possibility of 
precisely define its perturbative expansion. For this reason, and also 
because the pole mass can be only defined in the heavy quark case, a short 
distance definition of the quark mass, as the $\MSbar$ or the invariant quark
masses, should be always preferred.

Despite the great theoretical effort which has been spent in the last few years 
to determine the values of quark masses, these quantities still remain 
among the least well known parameters in the Standard Model. This is expecially 
true in the light quark sector, where the relative uncertainties on the up, 
down and strange quark masses are estimated to be as large as $50\%$ \cite{pdg}.

Lattice QCD is in principle able to predict the mass of any quark by fixing,
to its experimental value, the mass of a hadron containing a quark with the
same flavour \cite{lub}. The quark mass that is directly determined in 
lattice simulations is the (short distance) bare lattice quark mass $m(a)$, 
where $a$ is the lattice spacing, the inverse of the UV cut-off. The 
conversion factor relating $m(a)$ to a continuum renormalized mass, $m(\mu)$,
only depends on typical scales of the order of $\mu \sim a^{-1} \sim 2-4$ 
GeV. In this respect lattice QCD is unique, since QCD sum rules calculations
of quark masses have to work at much smaller scales, where higher-order 
corrections and/or non-perturbative effects \cite{istan} may be rather large.

It has been observed that perturbative calculations, within the lattice
regularization, suffer from large higher-order corrections \cite{lep-mac}. 
This is due to the presence of tadpole-like diagrams, which are absent in 
continuum perturbation theory. Such corrections might then introduce a large 
uncertainty in the calculation of the renormalization constant relating 
the bare lattice quark mass $m(a)$ to the renormalized mass $m(\mu)$ in a 
continuum scheme. This constant is only known at one-loop order 
\cite{rwil,rclov}, and its NLO renormalization group-improved version has 
been derived in ref.~\cite{all94}.

The uncertainties related to the perturbative calculation of the lattice
renormalization constants can be completely avoided by using the 
non-perturbative renormalization program proposed in ref.~\cite{ril}. 
Within this approach, the renormalization constants of lattice bare correlation 
functions, or couplings and masses, are computed in a non-perturbative way 
in the so called ``Regularization Invariant" ($\RI$) scheme. When required, 
the connection between the renormalized quantities in the $\RI$ scheme and 
those defined in any other scheme, like the $\MSbar$ one, can be then 
computed in continuum perturbation theory.

Recently, an extensive lattice calculations of light quark masses, with 
non-perturbatively determined renormalization constants in the $\RI$ scheme,
has been performed in ref.~\cite{ggrt}.%
\footnote{An explorative study can be found in ref.~\cite{all94}.}
The relation between the quark mass in the $\RI$ and $\MSbar$ schemes has 
been computed at one loop in ref.~\cite{ril}. This result, combined with the 
two-loop expansion of the QCD $\beta$-function and the quark mass anomalous 
dimension, has been then converted into a NLO renormalization group-improved 
determination of the quark masses. 

At present, both the $\beta $-function and the mass anomalous dimension in 
the $\MSbar$ scheme have been computed up to four loops, in refs.~\cite{b4} 
and \cite{g4a,g4b} respectively. Thus, the two-loop conversion factor 
$R_m$, relating the quark masses in the $\RI$ and $\MSbar$ schemes, is the 
only missing ingredient necessary to convert the non-perturbative lattice 
determinations of the quark masses to their $\MSbar$ counterparts at the NNLO. 
Such a calculation is the main result of this paper. The factor $R_m$ is given
in eq.~(\ref{eq:rm-lan}) for $N_c=3$ and in the Landau gauge, which is the
gauge usually adopted in $\RI$-scheme lattice calculations. We find that the
size of the NNLO corrections is quite large, of the order of $7\%$ at a 
typical scale $\mu \sim 2$ GeV, and comparable with the NLO contribution.

The paper is organized as follows. In sec.~\ref{sec:rs}, we discuss the
renormalization conditions for the quark mass, by referring in particular to
the $\MSbar$ and $\RI$ schemes. The perturbative expressions for the relevant 
renormalization constants and the ratio $R_m$ are given in sec.~\ref{sec:rc}. 
In sec.~\ref{sec:evol} we discuss the evolution of the running quark mass, we 
introduce the invariant quark mass and express the result of our calculation
as a relation between the invariant mass and the mass in the $\RI$ scheme.

\section{Quark mass renormalization schemes}
\label{sec:rs} 

The quark mass renormalization constant, $Z_m\left( \mu \right)$, relates the 
bare quark mass $m_0$ to the renormalized mass $m\left( \mu \right)$, defined
at a scale $\mu$ in a given scheme: 
\begin{equation}
\label{eq:zm}
m\left( \mu \right) =Z_m^{-1}\left( \mu \right) m_0 
\end{equation}
This definition provides the relation between the values of quark masses in
different renormalization schemes. By considering for definitiveness the 
$\MSbar$ and $\RI$ schemes, one finds: 
\begin{equation}
\label{eq:ratio}
R_m \left( \mu \right) = \frac{m^{\MSbar}\left( \mu \right)}{m^{\RI}\left( 
\mu \right)} = \frac{Z_m^{\RI}\left( \mu \right)}{Z_m ^{\MSbar}\left( \mu 
\right) }
\end{equation}
Even though $Z_m^{\MSbar}\left( \mu \right) $ and $Z_m^{\RI}\left(\mu 
\right)$ are separately divergent in the infinite cut-off limit, the ratio 
$R_m$ is finite, being equal to the ratio of renormalized quark masses in 
different schemes.

The quark mass renormalization can be conveniently defined through a
renormalization condition for the inverse bare quark propagator in momentum
space, $S_0^{-1}(p)$. In perturbation theory, $S_0^{-1}(p)$ has the form: 
\begin{equation}
\label{eq:s0inv}
i\,S_0^{-1}(p)=\slashchar{p}\,\Sigma _1(p)-m_0\,\Sigma _2(p) 
\end{equation}
where $\Sigma _{1,2}$ are (bare) scalar quantities which depend also on the
cut-off and the bare coupling constant, mass and gauge parameter, $\alpha 
_0$, $m_0$ and $\xi_0$. Once expressed in terms of renormalized parameters, 
$\as$, $m$ and $\xi $, the renormalization of the quark propagator still 
requires the renormalization of the quark field itself. Thus, the 
renormalized propagator can be written as: 
\begin{equation}
\label{eq:srinv}
i\,S^{-1}(p)=Z_q\left[ \slashchar{p}\,\overline{\Sigma } _1(p)-Z_mm\,
\overline{\Sigma }_2(p)\right] 
\end{equation}
where $\overline{\Sigma }_{1,2}$ are the bare functions $\Sigma _{1,2}$
expressed in terms of renormalized parameters.

Different renormalization prescriptions on $\overline{\Sigma }_{1,2}$ define
the expressions of the quark field and mass renormalization constants, $Z_q$
and $Z_m$, in the corresponding schemes. In perturbation theory, these
prescriptions can be conveniently expressed by giving the (finite)
coefficients entering in the perturbative expansions of the renormalized
quantities. Thus, for instance, for $\overline{\Sigma }_{1,2}$, up to ${\cal 
O}\left( \as^2\right)$, a renormalization scheme can be defined by: 
\begin{equation}
\label{eq:rs12} 
\begin{array}{l}
\lim \limits_{m\rightarrow 0} \dfrac 1{48} Z_q {\rm Tr}\left[ \gamma _\mu 
\dfrac{\partial \left( \slashchar{p}\,\overline{\Sigma }_1\left( p \right) 
\right) }{\partial p_\mu }\right] 
\Sb p^2=-\mu ^2 \\ \xi =\xi ^{*} \quad \endSb
= 1+\dfrac{\as}{\left( 4\pi \right) }\ r_1+ \dfrac{\as^2}{\left( 4\pi 
\right) ^2}\ s_1 + \ldots \\ 
\lim \limits_{m\rightarrow 0} \dfrac 1{12} Z_q Z_m {\rm Tr} \left[ 
\overline{\Sigma }_2\left( p \right) \right] 
\Sb p^2=-\mu ^2 \\ \xi =\xi ^{*} \quad \endSb
= 1 + \dfrac{\as}{\left( 4\pi \right) }\ r_2+\dfrac{
\as^2}{\left( 4\pi \right) ^2}\ s_2 + \ldots 
\end{array}
\end{equation}
where traces are taken over the color and spin indices.

The popular $\mbox{MS}$ scheme \cite{ms} amounts to require that, in
dimensional regularization, the coefficients of the perturbative expansion of
the renormalization constants only contain singular terms in $1/\epsilon $,
where $D=4-2\epsilon$ is the space-time dimension. Here we consider its
standard modification, the $\MSbar$ scheme \cite{msbar}. We choose to 
renormalize the gauge coupling constant $\as$ and the gauge parameter $\xi$ 
always in this scheme, independently on the renormalization condition adopted 
for the quark field and mass. A notable feature of the $\MSbar$ scheme is 
that the renormalization constants of gauge-invariant operators and 
parameters, like the quark mass itself, are gauge independent. Consequently, 
the renormalized quark mass, in the $\MSbar$ scheme, is gauge independent.

One of the advantage of the $\RI$ scheme is that the corresponding
renormalization prescriptions can be stated in a non-perturbative way. In the 
case of $\overline{\Sigma }_{1,2}$, for instance, one simply requires the 
left-hand sides of eqs.~(\ref{eq:rs12}) to be equal to unity. In perturbation
theory, this implies the vanishing of all the coefficients $r_i$ and $s_i$. 
The renormalization conditions in the $\RI$ scheme clearly depend on the 
choice of the gauge parameter $\xi^{*}$, and different choices define in fact 
different $\RI$ renormalization schemes. Throughout this paper we will present
results for the $\RI$ scheme in a generic covariant gauge, even though a common
choice in this context is the Landau gauge, corresponding to $\xi^{*}=0$. 
Indeed, this choice is particularly convenient to implement in non-perturbative
calculations of lattice QCD.

In numerical lattice simulations, the quark field renormalization condition 
of eq.~(\ref{eq:rs12}) is not easy to implement, because of the presence of 
the continuum derivative. A common practice is to define a different quark
field renormalization through the condition:
\be
\lim \limits_{m\rightarrow 0}\dfrac 1{12}Z_q^\prime 
{\rm Tr}\left[\overline{\Sigma }_1\left( p \right) \right]
\Sb p^2=-\mu ^2 \\ \xi =\xi ^{*} \quad \endSb = 1
\label{eq:zqp}
\ee
In the Landau gauge, the difference between the two definitions appears at
order $O(\as^2)$, and it must be taken into account at the NNLO accuracy.
For this reason, we have also computed the ratio of the two relevant 
renormalization constants and the result will be given in the next section.

In the calculation of $\overline{\Sigma }_{1,2}$ an important check is
provided by the vector chiral Ward identity, which relates these functions
to the amputated Green functions of the vector current and scalar density
respectively between external quark states, $\Gamma _{V_\mu }(p)$ and 
$\Gamma _S(p)$. In particular, in the chiral limit: 
\begin{equation}
\label{eq:wid}
\left[ \dfrac{\partial \left( \slashchar{p}\,\overline{\Sigma }
_1\left( p\right) \right) }{\partial p_\mu }=\Gamma _{V_\mu }(p)\right]
_{m=0}\qquad ,\qquad \left[ \overline{\Sigma }_2\left( p\right) =\Gamma
_S(p)\right] _{m=0} 
\end{equation}
We have verified that the above equations are indeed satisfied at the level
of bare Green functions. By requiring the Ward identities to be satisfied
also by renormalized quantities, one obtains the relations: 
\begin{equation}
\label{eq:zvzs}
Z_V=1\qquad ,\qquad Z_m=Z_S^{-1} 
\end{equation}
among the renormalization constants in a generic scheme. In the $\MSbar -$NDR 
scheme, the vector and axial Ward identities are unaffected by the minimal 
subtraction procedure. Thus, in this scheme, eqs.~(\ref{eq:zvzs}) are 
automatically satisfied. In the $\RI$ scheme, in order to satisfy the Ward 
identities, the following renormalization prescriptions on $\Gamma _{V_\mu }$ 
and $\Gamma _S$ must be consistently applied: 
\begin{equation}
\label{eq:rivs} 
\begin{array}{l}
\lim \limits_{m\rightarrow 0} \dfrac 1{48} Z_q^{\RI} \left( Z_V^{\RI}\right) 
^{-1} {\rm Tr} \left[ \gamma _\mu \Gamma _{V_\mu }(p)\right]
\Sb p^2=-\mu ^2 \\ \xi =\xi ^{*} \quad \endSb =1 \\ 
\lim \limits_{m\rightarrow 0} \dfrac 1{12} Z_q^{\RI}\left(Z_S^{\RI}\right) 
^{-1} {\rm Tr} \left[ \Gamma _S(p)\right] 
\Sb p^2=-\mu ^2 \\ \xi =\xi ^{*} \quad \endSb =1 
\end{array}
\end{equation}
The above conditions can be also applied at finite values of the quark mass,
provided a sufficiently large value of the renormalization scale is chosen,
$\mu^2 \gg m^2$. 

\section{Results for the renormalization constants}
\label{sec:rc} 

In order to present the results of our calculation, we expand a generic 
renormalization constant $Z$ as a series in the strong coupling constant: 
\begin{equation}
\label{eq:Zexp}
Z=1+\frac{\as}{4\pi }Z^{(1)}+\frac{\as^2}{(4\pi)^2}Z^{(2)}+\ldots
\end{equation}
Subsequently, each coefficient $Z^{(i)}$ is expanded in inverse powers of 
$\epsilon$: 
\begin{equation}
\label{eq:Zpoleexp}
Z^{(i)}=\sum_{j=0}^i\left( \frac 1\epsilon \right)
^jZ_j^{(i)}
\end{equation}

By definition, all the coefficients $Z_0^{(i)}$ vanish in the $\MSbar$ scheme. 
In addition, by requiring that the ratio of renormalization constants
in different schemes, namely $Z^{\RI}/Z^{\MSbar}$, is finite, one finds a set 
of useful identities between the $Z_j^{(i)}$s in the two schemes. These 
observations, combined together, imply the following relations: 
\begin{equation}
\label{eq:rel2sc} 
\begin{array}{l}
\left( Z^{\MSbar}\right) _1^{(1)}=\left( Z^{\RI}\right) _1^{(1)} \\ 
\left( Z^{\MSbar}\right) _2^{(2)}=\left( Z^{\RI}\right) _2^{(2)} \\ 
\left( Z^{\MSbar}\right) _1^{(2)}=\left( Z^{\RI}\right) _1^{(2)} - 
\left( Z^{\RI}\right) _1^{(1)}\left( Z^{\RI}\right) _0^{(1)} 
\end{array}
\end{equation}
Thus, the renormalization constant in the $\MSbar$ scheme can be completely 
derived from its counterpart in the $\RI$ scheme (but not vice-versa). One 
also finds that the perturbative expansion of the ratio $Z^{\RI}/Z^{\MSbar}$ 
is simply given, up to two-loop order, by the finite coefficient of 
perturbative expansion of $Z^{\RI}$: 
\begin{equation}
\label{eq:rm}
R_m = \frac{Z_m^{\RI}}{Z_m^{\MSbar}} = 
1+ \dfrac{\as}{\left( 4\pi \right) }\left( Z_m^{\RI}\right)_0^{(1)}
+\dfrac{\as^2}{\left( 4\pi \right) ^2}\left( Z_m^{\RI}\right) _0^{(2)} 
+ \ldots
\end{equation}

\begin{figure}
\begin{center}
\epsfxsize=\textwidth
\leavevmode\epsffile{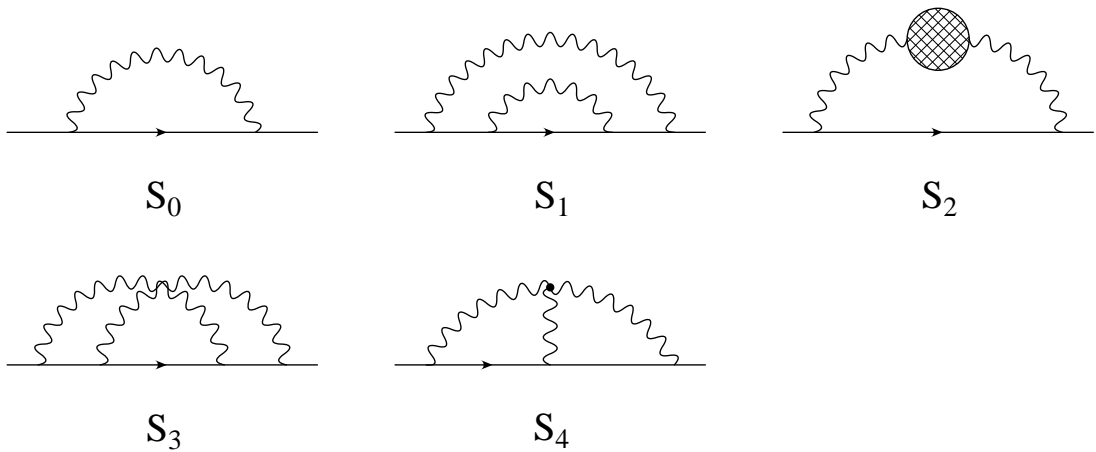}
\caption[]{\it One- and two-loop diagrams for the quark propagator.}
\label{fig:fd}
\end{center}
\end{figure}

We have calculated the quark field and mass renormalization constants, $Z_q$
and $Z_m$, in the $\MSbar$ and $\RI$ schemes. Adopting na\"\i ve dimensional 
regularization, in a generic covariant gauge, we have computed the Feynman 
diagrams shown in fig.~\ref{fig:fd}. Since the corresponding amplitudes can
be expanded up to linear terms in the bare quark mass, this only requires 
the evaluation of one- and two-loop p-integrals, which we have computed by
using the method of integration by parts of ref.~\cite{pint}. The Ward 
identities of eqs.~(\ref{eq:wid}) have been used to check the correctness of 
our calculation. 
For the quark field renormalization constant, in the $\RI$ scheme, we find: 
\begin{eqnarray}
\label{eq:zqri}
\left( Z^{\RI}_q \right)^{(1)}_{1} & = & 
  {{\left(N_c^2 - 1 \right) }\over {2\,N_c}} \, 
  \left( - \xi \right) \nonumber\\
\left( Z^{\RI}_q \right)^{(1)}_{0} & = & 
  {{\left(N_c^2 - 1 \right) }\over {4\,N_c}} \, 
  \left( - \xi \right) \nonumber\\
\left( Z^{\RI}_q \right)^{(2)}_{2} & = & 
  {{\left(N_c^2 - 1 \right) }\over {8\,N_c^2}} \, \xi \, 
 \left( -\xi + 3\,N_c^2 + 2\,\xi\,N_c^2 \right)  \\
\left( Z^{\RI}_q \right)^{(2)}_{1} & = & 
 {{\left(N_c^2 - 1 \right) }\over {16\,N_c^2}} \,
 \left( -3 - 2\,\xi^2 - 22\,N_c^2 - 8\,\xi\,N_c^2 + 
 {\xi^2}\,N_c^2 + 4\,N_c\,n_f \right) \nonumber\\
\left( Z^{\RI}_q \right)^{(2)}_{0} & = & 
 {{\left( N_c^2 - 1 \right) }\over {32\,N_c^2}} \,
 \left( 1 - 6\, \xi^2 - 115\,N_c^2 - 76\,\xi\,N_c^2 - 
 6\, \xi^2\,N_c^2 \right. \nonumber \\
 & \phantom{x} & \phantom{xxxxxxxx} \left.
  + 48\,\xi\,N_c^2\zeta_3 + 48\,N_c^2\zeta_3 + 20\,N_c\,n_f \right) \nonumber
\end{eqnarray}
where $\zeta$ is the Riemann zeta function and $\zeta _3=1.20206\ldots$.
The quark mass renormalization constant, in the same scheme, is given by:
\begin{eqnarray}
\label{eq:zmri}
\left( Z^{\RI}_m \right)^{(1)}_{1} & = & 
  {{\left(N_c^2 - 1 \right) }\over {2\,N_c}} \,
  \left( - 3 \right) \nonumber\\
\left( Z^{\RI}_m \right)^{(1)}_{0} & = & 
  {{\left(N_c^2 - 1 \right) }\over {4\,N_c}} \, 
  \left( - 8 - 3\,\xi \right) \nonumber\\
\left( Z^{\RI}_m \right)^{(2)}_{2} & = & 
  {{\left(N_c^2 - 1 \right) }\over {8\,N_c^2}}  \,
  \left( - 9 + 31\,N_c^2 - 4\,N_c\,n_f \right) \\
\left( Z^{\RI}_m \right)^{(2)}_{1} & = & 
  {{\left(N_c^2 - 1 \right) }\over {48\,N_c^2}} \,
  \left( -135 - 54\,\xi - 59\,N_c^2 + 54\,\xi\,N_c^2 + 
  20\,N_c\,n_f \right)  \nonumber\\
\left( Z^{\RI}_m \right)^{(2)}_{0} & = & 
  {{\left(N_c^2 -1 \right) }\over {96\,N_c^2}} \,
  \left( -75 - 144\,\xi - 36\, \xi^2 - 2645\,N_c^2
- 108\,\xi\,N_c^2 \right.  \nonumber\\
& \phantom{x} & \phantom{xxxxxxxx}
 \left. - 18\, \xi^2\,N_c^2
  + 288\,\zeta_3 + 576\,N_c^2\,\zeta_3 + 356\,N_c\,n_f  \right) \nonumber
\end{eqnarray}
The corresponding values of renormalization constants in the $\MSbar$ scheme 
can be obtained from eqs.~(\ref{eq:zqri}) and (\ref{eq:zmri}) by using 
eq.~(\ref{eq:rel2sc}). The result for $Z_m^{\MSbar}$ is in agreement with
the original calculation of ref.~\cite{tarr}. Notice that, as expected, 
$Z_m^{\MSbar}$ turns out to be gauge-independent.

We have also calculated the ratio $\Delta_q$ between the quark field 
renormalization constants $Z_q^{\RI}$ and $Z_q^\prime$, the latter being 
defined in eq.~(\ref{eq:zqp}). The result, expanded in series of the
strong coupling constant, is:
\begin{equation}
\label{eq:deltaq1}\Delta _q=\frac{Z_q^{\RI}}{Z_q^\prime}=
1+\dfrac{\as}{\left( 4\pi \right) }\Delta
_q^{(1)}+\dfrac{\as^2}{\left( 4\pi \right) ^2}\Delta _q^{(2)}+\ldots
\end{equation}
where we find:
\begin{eqnarray}
\label{eq:deltaq2}
\Delta^{(1)}_q & = & 
  {{\left(N_c^2 - 1 \right) }\over {4\,N_c}} \, \xi \\
\Delta^{(2)}_q & = & 
 {{\left( N_c^2 - 1 \right) }\over {16\,N_c^2}} \,
 \left(3 - \xi^2 + 22 N_c^2 + 14\xi N_c^2 + 4\xi^2 N_c^2
  - 4 N_c n_f \right) \nonumber
\end{eqnarray}
Notice that, in the Landau gauge $\left( \xi =0\right)$, at a typical scale 
$\mu \sim 2$ GeV, $\Delta_q$ represents a tiny correction, smaller than $1\%$.

Finally, we discuss the results for the ratio $R_m$, which provides the 
relation between the quark mass in the $\MSbar$ and $\RI$ schemes at a fixed 
scale $\mu$. This ratio is obtained by substituting the values of $\left(Z_m
^{\RI}\right)_0^{(1)}$ and $\left( Z_m^{\RI}\right) _0^{(2)}$ in 
eq.~(\ref{eq:rm}). For convenience, we present here its numerical expression
as obtained for $N_c=3$ in the Landau gauge: 
\begin{equation}
\label{eq:rm-lan}
R_m^{LAN}\left( \mu \right) = 1- \frac{16}3 \dfrac{\as\left( \mu \right)}{
\left( 4\pi \right)}-\left( \frac{1990}9 - \frac{152}{3}\zeta_3
-\frac{89}9n_f\right) \dfrac{\as^2
\left( \mu \right) }{\left( 4\pi \right) ^2} 
\end{equation}
Eq.~(\ref{eq:rm-lan}) is the main result of this paper. Numerically, one
finds that the size of the NNLO contribution to $R_m^{LAN}$, at a scale 
$\mu =$2 GeV $\left( n_f=4\right) $, is about $7\%$. This represents a 
significant correction, comparable to the NLO contribution.

\section{Quark mass evolution and the invariant quark mass}
\label{sec:evol}
 
In this section we discuss the NNLO relations between quark masses in 
different renormalization schemes and at different renormalization scales. 
This discussion also provides us with the opportunity to introduce the 
so called renormalization group-invariant quark mass, $\widehat{m}$, and its 
perturbative relation with the quark mass in the $\RI$ scheme.

In a generic renormalization scheme, the renormalized quark mass obeys the
following renormalization group equation: 
\begin{equation}
\label{eq:rge}
\left[ \mu ^2\frac d{d\mu ^2}+\frac{\gamma _m}2\right] m\lp \mu \rp =
\left[ \beta \left( \as\right) \frac \partial {\partial \as}+\beta _\xi 
\left( \as\right) \xi \frac \partial {\partial \xi }+\frac{\gamma _m} 
2\right] m \lp \mu \rp =0 
\end{equation}
Both the $\beta $-function and the mass anomalous dimension in the $\MSbar$ 
scheme have been computed up to four loops in \cite{b4} and \cite{g4a,g4b} 
respectively. In the NNLO approximation we are considering here, we only need 
the corresponding expansions up to three loops. The QCD $\beta $-function is 
given by: 
\begin{eqnarray}
\frac{\beta (\as)}{4\pi} & = & \mu ^2 \frac{d}{d\mu ^2} \lp \frac{\as}{4\pi} 
\rp \, = \, - \sum_{i=0}^\infty \beta _i\left( \frac{\as}{4\pi } \right) 
^{i+2} \nn \\
\beta _0 & = & \frac{11}{3}N_c - \frac{2}{3} n_f \\
\beta _1 & = & \frac{34}3N_c^2-\frac{10}3N_cn_f-\frac{(N_c^2-1)}{N_c}n_f 
\nonumber \\
\beta _2^{\MSbar} & = & \frac{2857}{54}N_c^3+\frac{\left(N_c^2-1\right) ^2}{
4 N_c^2}n_f-\frac{205}{36}\left( N_c^2-1\right) n_f 
\nonumber \\
&\phantom{x}&\;-\frac{1415}{54}N_c^2n_f+\frac{11}{18}\frac{\left(N_c^2-
1\right) }{N_c}n_f^2+\frac{79}{54}N_cn_f^2 \nonumber 
\label{eq:beta}
\end{eqnarray}
The mass anomalous dimension in the $\MSbar$ scheme is: 
\begin{eqnarray}
\gamma (\as) &=& 2Z^{-1}\mu ^2\frac d{d\mu ^2}Z=
\sum_{i=0}^\infty\gamma^{(i)}\left( \frac{\as}{4\pi }\right) ^{i+1}
\nonumber \\
\gamma_m^{(0)} &=&3\frac{N_c^2-1}{N_c}  \nonumber \\
\gamma_m^{(1)} &=&\frac{N_c^2-1}{N_c^2}\left( -\frac 34+%
\frac{203}{12}N_c^2-\frac 53N_cn_f\right)  \\
\gamma_m^{(2)} &=&\frac{N_c^2-1}{N_c^3}\left[ \frac{129}8-%
\frac{129}8N_c^2+\frac{11413}{108}N_c^4\right.   \nonumber \\
&\phantom{x}&\;\left. +n_f\left( \frac{23}2N_c-\frac{1177}{54}%
N_c^3-12N_c\zeta _3-12N_c^3\zeta _3\right) -\frac{35}{27}N_c^2n_f^2\right] 
\nonumber  \label{eq:gam}
\end{eqnarray}
The evolution of the renormalized gauge parameters is given by:
\begin{eqnarray}
\beta_\xi (\as) & = & \frac{\mu ^2}{\xi} \frac{d \xi}{d\mu ^2}\, = \, 
- \sum_{i=1}^\infty \beta _\xi^{(i)} \left( \frac{\as}{4\pi } \right) 
^i \nn \\
\beta _\xi^{(0)} & = & -\frac{N_c}{2} \lp \frac{13}{3} - \xi \rp + 
\frac{2}{3} n_f
\label{eq:betacsi}
\end{eqnarray}

For completeness, we also present the coefficients of the perturbative
expansion of the quark field anomalous dimension, up to the NLO. Using the 
results of eq.~(\ref{eq:zqri}), we derive, in the $\MSbar$ scheme and in a 
generic covariant gauge:
\bea
\gamma_q^{(0)}&=& \frac{N_c^2-1}{N_c}\xi \nn \\
\gamma_q^{(1)}&=&\frac{N_c^2-1}{4N_c^2}\left( 3+22N_c^2+8\xi N_c^2
+\xi^2 N_c^2-4N_c n_f \right)
\label{eq:wfad}
\eea

The evolution of the quark mass is determined by eq.~(\ref{eq:rge}).
The solution is particularly simple in the $\MSbar$ scheme, where the 
renormalized quark mass is gauge independent. In this case, it can be 
expressed in the form: 
\begin{equation}
\label{eq:msol}
m^{\MSbar}\left( \mu \right) =\frac{c^{\MSbar}\left(\mu \right) }{c
^{\MSbar}\left(\mu _0 \right) }\ m^{\MSbar}\left( \mu _0\right) 
\end{equation}
where:
\begin{eqnarray}
c^{\MSbar} \left( \mu \right) &=&\as \left( \mu \right)^{\overline{\gamma }
_0}\left\{ 1 +\frac{\as}{4\pi } \left( \overline{\gamma }_1-\overline{\beta }
_1 \overline{\gamma }_0\right) \right.  \\
&+&\left. \frac 12\left( \frac{\as \left( \mu \right)}{4\pi }\right) ^2\left[
\left( \overline{\gamma }_1-\overline{\beta }_1\overline{\gamma }_0\right) 
^2+ \overline{\gamma }_2+\overline{\beta }_1^2\overline{\gamma }_0-
\overline{\beta }_1\overline{\gamma }_1-\overline{\beta }_2\overline{\gamma 
}_0 \right] \right\} \nonumber  
\label{eq:calfa}
\end{eqnarray}
with $\overline{\beta }_i=\beta _i/\beta _0$ and $\overline{\gamma }%
_i=\gamma_m^{(i)}/\left( 2\beta _0\right) $.

By using eq.~(\ref{eq:ratio}), we can convert eq.~(\ref{eq:msol}) into a
relation between the $\RI$ quark mass at a scale $\mu _0$ and the $\MSbar$ 
one at a different scale $\mu $: 
\begin{equation}
\label{eq:m2}
m^{\MSbar}\left( \mu \right) =\frac{c^{\MSbar} \left( \mu\right)}{c^{\MSbar} 
\left( \mu _0 \right) }\ \frac{Z^{\RI}\left( \mu _0\right) }{Z^{\MSbar}\left(
\mu _0\right) }\ m^{\RI}\left( \mu _0\right) 
\end{equation}
This formula can be now easily interpreted by observing that the function: 
\begin{equation}
\label{eq:cri}
c^{\RI}\left( \mu \right) =\frac{Z^{\MSbar}\left( \mu \right) 
}{Z^{\RI}\left( \mu \right) } \ c^{\MSbar} \left( \mu \right) 
\end{equation}
is the solution of eq.~(\ref{eq:rge}) in the $\RI$ scheme. So we can
conventionally define a renormalization group-invariant mass as: 
\begin{equation}
\label{eq:mhat}
\widehat{m}=\frac{m^{\MSbar}\left( \mu \right)}{c^{\MSbar}\left( \mu \right)
}=\frac{m^{\RI}\left( \mu _0\right) }{c^{\RI}\left( \mu_0 \right) }
\end{equation}
The mass $\widehat{m}$ is a short-distance quantity, which is both scale and 
scheme independent. It can be conveniently used in phenomenological 
applications involving quark masses, in alternative to the $\MSbar$ quark 
mass.

The relation between $\widehat{m}$ and the $\MSbar$ quark mass has been 
computed in ref.~\cite{g4a}. For $n_f=3,4,5$, the result has the form: 
\begin{eqnarray}
\widehat{m}^{\left( 3\right)} &=& \as\left(\mu\right)^{-4/9}
\left[ 1 - \frac{\as\left(\mu\right)}{4\pi}\frac{290}{81} - 
\frac{\as^2 \left(\mu\right)}{\left( 4\pi \right)^2 }
\left( \frac{259943}{13122} -\frac{80}{9} \zeta _3 \right) \right]
m^{\MSbar}\left(\mu\right) 
\\
\widehat{m}^{\left( 4\right)} &=& \as\left(\mu\right)^{-12/25}
\left[ 1 - \frac{\as\left(\mu\right)}{4\pi}\frac{7606}{1875} - 
\frac{\as^2 \left(\mu\right)}{\left( 4\pi \right)^2 }
\left( \frac{446305267}{21093750} - \frac{64}{5}\zeta _3 \right)\right]
m^{\MSbar} \left(\mu\right)  \nonumber 
\\
\widehat{m}^{\left( 5\right)} &=& \as\left(\mu\right)^{-12/23}
\left[ 1 - \frac{\as\left(\mu\right)}{4\pi}\frac{7462}{1587} - 
\frac{\as^2 \left(\mu\right)}{\left( 4\pi \right)^2 }
\left( \frac{344665349}{15111414} - \frac{400}{23}\zeta _3 \right)\right]
m^{\MSbar} \left(\mu\right) \nonumber
\end{eqnarray}
We can use eq.~(\ref{eq:cri}) to state our results of 
eqs.~(\ref{eq:rm})-(\ref{eq:rm-lan}) as a relation between the renormalization 
group-invariant mass $\widehat{m}$ and the mass $m^{\RI}$: 
\begin{eqnarray}
\widehat{m}^{\left( 3\right)} &=& \as\left(\mu\right)^{-4/9}
\left[1 - \frac{\as\left(\mu\right)}{4\pi}\left(\frac{722}{81}+2\xi \right) 
\right. \nonumber\\
&\phantom{x}& \left. - \frac{\as^2 \left(\mu\right)}{\left( 4\pi \right)^2 }
\left( \frac{2521517}{13122} -\frac{536}{9}\zeta _3
+\frac{257}{81}\xi + \frac{11}{6}\xi^2 \right) \right] 
m^{\RI}\left(\mu\right)  \nonumber \\
\widehat{m}^{\left( 4\right)} &=& \as\left(\mu\right)^{-12/25}
\left[ 1 - \frac{\as\left(\mu\right)}{4\pi}\left( \frac{17606}{1875} +2\xi 
\right) \right. \\
&\phantom{x}& \left. - \frac{\as^2 \left(\mu\right)}{\left( 4\pi \right)^2 }
\left( \frac{3819632767}{21093750}-\frac{952}{15}\zeta _3 
+\frac{4163}{1875}\xi +\frac{11}{6}\xi^2 \right) \right] 
m^{\RI}\left(\mu\right)  \nonumber \\
\widehat{m}^{\left( 5\right)} &=& \as\left(\mu\right)^{-12/23}
\left[ 1 - \frac{\as\left(\mu\right)}{4\pi}\left( \frac{15926}{1587} +2\xi 
\right) \right.  \nonumber \\
&\phantom{x}& \left. - \frac{\as^2 \left(\mu\right)}{\left( 4\pi \right)^2 }
\left( \frac{2559841211}{15111414}-\frac{4696}{69}\zeta _3
+ \frac{1475}{1587}\xi + \frac{11}{6}\xi^2 \right) \right] 
m^{\RI}\left(\mu\right)  \nonumber
\end{eqnarray}
We note that in the $\RI$ scheme both $m^{\RI}$ and $c^{\RI}$ depend on the 
gauge, but the dependence cancels out in the ratio $\widehat{m}$.

\section*{Acknowledgements}
We warmly thank G.~Martinelli for many interesting discussions. We 
acknowledge the M.U.R.S.T. and the INFN for partial support.


\begin{thebibliography}{9}

\bibitem{g4a} J.A.M.~Vermaseren, S.A.~Larin and T.~van Ritbergen,
Phys.~Lett. B405 (1997) 327, {\tt hep-ph/9703284}.

\bibitem{pole} N.~Gray, D.J.~Broadhurst, W.~Grafe and K.~Schilcher,
Z.~Phys. C48 (1990) 673.

\bibitem{renorm1} I.I.~Bigi, M.A.~Shifman, N.G.~Uraltsev and A.I.~Vainshtein,
Phys.~Rev. D50 (1994) 2234, {\tt hep-ph/9402360}.

\bibitem{renorm2} M.~Beneke and V.M.~Braun, Nucl.~Phys. B426 (1994), 301,
{\tt hep-ph/9402364}.

\bibitem{pdg} {\em The Particle Data Group}, R.M.~Barnett {\em et al.}, 
Phys.~Rev. D54 (1996) 1.

\bibitem{lub} For a review of recent results, see V.~Lubicz, talk given at 
the APCTP-ICTP Joint International Conference (AIJIC 97), Seoul, Korea, 
26-30 May 1997, {\tt hep-ph/9707374}.

\bibitem{istan} E.~Gabrielli and P.~Nason, Phys.~Lett. B313 (1993) 430.

\bibitem{lep-mac} G.P.~Lepage and P.B.~Mackenzie, Phys.~Rev. D48 (1993),
2250, \\ {\tt hep-lat/9209022}. 

\bibitem{rwil} A.~Gonzales Arroyo, F.J.~Yndurain and G.~Martinelli,
Phys.~Lett. B117 (1982) 437. \\
G.~Martinelli and Y.C.~Zhang, Phys.~Lett. B123 (1983) 433. \\ 
H.W.~Hamber and C.M.~Wu, Phys.~Lett. B133 (1983) 351.

\bibitem{rclov} E.~Gabrielli, G.~Martinelli, C.~Pittori, G.~Heatlie and
C.T.~Sachrajda, Nucl.~Phys. B362 (1991) 475. \\
A.~Borrelli, C.~Pittori, R.~Frezzotti and E.~Gabrielli, Nucl.~Phys. B409 
(1993) 382. 

\bibitem{all94} C.R.~Allton, C.T.~Sachrajda, V.~Lubicz, L.~Maiani and
G.~Martinelli, Nucl.~Phys. B349 (1991) 598. 

\bibitem{ril}  G.~Martinelli, C.~Pittori, C.T.~Sachrajda, M.~Testa and 
A.~Vladikas, Nucl.~Phys. B445 (1995) 81, {\tt hep-lat/9411010}.

\bibitem{ggrt} V.~Gimenez, L.~Giusti, F.~Rapuano and M.~Talevi,
EDINBURGH-97-15, Jan 1998, {\tt  hep-lat/9801028}.

\bibitem{b4}  T.~van Ritbergen, J.A.M.~Vermaseren and S.A.~Larin, 
Phys.~Lett. B400 (1997) 379, {\tt hep-ph/9701390}.

\bibitem{g4b}  K.G.~Chetyrkin, Phys.~Lett. B404 (1997) 161, 
{\tt hep-ph/9703278}.

\bibitem{ms} G.~'t Hooft, Nucl.~Phys. B61 (1973) 455.

\bibitem{msbar} W.A.~Bardeen, A.J.~Buras, D.W.~Duke and T.~Muta,
Phys.~Rev. D18 (1978) 3998.

\bibitem{pint} K.G.~Chetyrkin and F.V.~Tkachov, Nucl.~Phys. B192 (1981) 159. 

\bibitem{tarr} R.Tarrach, Nucl.~Phys. B183 (1981) 384. 

\end{thebibliography}
\end{document}